\documentclass[12pt]{article}

\usepackage{amssymb,amsmath}

\usepackage{graphicx}
\DeclareGraphicsExtensions{.pdf,.png,.jpg}

 \catcode`\@=11 \@addtoreset{equation}{section}\catcode`\@=12

\begin{document}

 \begin{center}

 \large \bf  Scale Factors of  Homogeneous Anisotropic Cosmological Models with Perfect Fluid 
  \end{center}

 \vspace{0.3truecm}

 \begin{center}

  K. K. Ernazarov

\vspace{0.3truecm}

 \it Institute of Gravitation and Cosmology,
 RUDN University, Miklukho-Maklaya ul. 6,
 Moscow, 117198, Russian Federation


\end{center}


The article is devoted to cosmology. It deals with homogeneous anisotropic cosmological models.  Scale factors have been evaluated for the multicomponent models with perfect fluid, taking account of  its expansion, rotation and shear. The cosmological fluid components are as follows: phantom matter, de Sitter vacuum, domain walls, strings, dust, radiation, perfect gas, stiff matter and ekpyrotic matter.  The scale factor dependences on the equation of state and kinematic invariants of the perfect fluid have been analysed. The paper  is aimed at finding exact solutions of Friedmann and Raychaudhuri equations for different combination of matter components and kinematic invariants.



\section{Introduction}

Cosmology is a division of astronomy studying the Universe as a whole. Cosmology is divided into observational and theoretical ones. The observational cosmology is based on extragalactic astronomy. The theoretical cosmology is formulated in the framework of general relativity. In the models of theoretical cosmology, the recession of galaxies is interpreted as an expansion of space-time itself, otherwise as the Universe expansion. The Universe is considered to be homogeneous and isotropic (Copernicus principle) on the scales exceeding 100 Mpc. Two models satisfy the Copernicus principle: the model proposed by Willem de Sitter  in 1917 and the model proposed by Alexander Friedmann in 1922.
	In the de Sitter model, the galaxies are considered to be test bodies moving along geodesics of space-time, whose curvature is constant and determined by a cosmological term in Einstein - Hilbert's equations, but not by masses. In Friedmann's model, space-time is determined by the masses of all galaxies, in whose gravitational field a test galaxy is moving. The metric of a homogeneous isotropic space-time depends on a scale factor (Universe's radius). The scale factor is found from Friedmann's equations for different matter components and their superposition.
	Homogeneous anisotropic space-time is described by Raychaudhuri's equation for kinematic invariants of cosmological fluid (expansion, divergence of acceleration, rotation and shear) with different equations of state. In the absence of anisotropy parameters,  Raychaudhuri's equation transforms into Friedmann's first equation.

\section{Problem statement}

The cosmological fluid motion is described, in general relativity, by Raychaudhuri's equation \cite{Hawking}, \cite{Fil'chenkov}
\begin{equation}
  \dot{\theta} + \frac{1}{3}{\theta}^2-A^i_{;i}+2({\sigma}^2-{\omega}^2)=-\frac{4{\pi}G}{c^2}(\epsilon+3p),
 \label{2.0}
\end{equation}
where $\theta$ is the expansion scalar, $A^i$  is the 4-acceleration, $\sigma$ is the shear scalar, $\omega$ is the rotation scalar, $\epsilon$ is the energy density, and p is the pressure.

The space-time metric reads
\begin{equation}
   ds^2=(N^2-N_aN^2)dt^2-2N_adtdx^{\alpha}-\gamma_{\alpha\beta}dx^{\alpha}dx^{\beta}  (\alpha, \beta=1,2,3)
   \label{2.1}
\end{equation}

where N is the lapse function, $N_a=cg_{0\alpha}$ is the shift function and the metric cross term $g_{0\alpha}$ satisfying the relation $g^{00}=1-g_{0\alpha}g^{0\alpha}$,$ \gamma_{11}\gamma_{22}\gamma_{33}=a^6$, $a(t)$ is the scale factor. The quantities $\theta$,
 $A^i$, $\sigma$, $\omega$ read

\begin{equation}
 \theta= u^i_{;i}, A^i=u^i_{;k}u^k,\\
\omega^2=\frac{1}{2}\omega_{ik}\omega^{ik}, \sigma^2=\frac{1}{2}\sigma_{ik}\sigma^{ik},
 \label{2.2}
\end{equation}
where $u_i$ is the 4-velocity. The orthogonality conditions have the form $u_i\omega_{ik}=0$, $u^i\sigma_{ik}=0$, $u^iA_i=0$.
From the conservation law $T^{ik}_{;k}=0$ for the energy momentum tensor
 \begin{equation}
 T^{ik}=(p+\epsilon)u^iu^k-pg^{ik}
       \label{2.5},
 \end{equation}
providing $\theta=3\dot{a}/a$, which is valid if $\sigma_{\alpha\beta}n^{\alpha}n^{\beta}=0$, where $n^{\alpha}n_{\alpha}=1$, we obtain an expression for the 4-acceleration divergence scalar as follows:
\begin{equation}
 A^i_{;i}=w(1-g^{00})(\dot{\theta}+\theta^2), g^{00}=const,
       \label{2.6},
\end{equation}
where $1-g^{00}$ characterizes a deviation from Friedmann's geometry. The equation of state is $p=w\epsilon$ where $w=const$.

\section{Homogeneous isotropic models}
The first observation important for cosmology was the discovery of a red shift in the spectral lines of distant galaxies. In models of theoretical cosmology, the recession of galaxies is interpreted as the expansion of the universe. At the same time from the Copernicus principle, it follows that there are no preferred  points and directions in the Universe, i.e. it is homogeneous and isotropic at scales larger than 100 Mpc. The Copernicus  principle is satisfied in two models: the model proposed by the Dutch astronomer Willem de Sitter in 1917 \cite{de Sitter}  and the model proposed by the Soviet mathematician Alexander Friedmann in 1922 \cite{Friedmann}.\\
In the de Sitter model, galaxies are considered to be test bodies moving along geodesics of space-time, whose the curvature is constant and determined by an additional term $\Lambda{g}_{ik}$, the so-called  cosmological term on the right-hand side of the Einstein-Hilbert equations, and not by the masses, thus $ T_{ik} = 0$. Finally, for this model, the Einstein-Hilbert equations take the form:
\begin{equation}
 R_{ik}-\frac{1}{2}Rg_{ik}=\Lambda{g}_{ik},
       \label{2.7},
\end{equation}
where $\Lambda$ is the cosmological constant.\\
In the Friedmann model, space-time is determined by the masses of all galaxies in the gravitational field of which the given test galaxy moves, thus $ T_{ik} \neq 0$.\\
For $g^{00} = 1$, and $\sigma = \omega = 0$, i.e., for homogeneous isotropic models, the Raychaudhuri equation (\ref{2.0})  is written as
\begin{equation}
 \ddot{a}=-\frac{4{\pi}G}{3c^2}(\epsilon +3p)a
       \label{2.8},
\end{equation}
which is a record of Newton's second law for cosmology, if we take into account the first law of thermodynamics
\begin{equation}
 a\dot{\epsilon}+3(p+\epsilon)\dot{a}=0
       \label{2.10}.
\end{equation}
Integrating equation (\ref{2.8}), we obtain the following equation
\begin{equation}
 \frac{{\dot{a}}^2}{2}-\frac{4{\pi}G\epsilon{a}^2}{3c^2}=-\frac{kc^2}{2}
       \label{2.9},
\end{equation}
In general, the energy density of a multicomponent medium has the form \cite{Fil'chenkov_02}:
\begin{equation}
 \epsilon=\epsilon_0\sum_{u}B_u{(\frac{r_0}{a}})^u,
       \label{2.11}
\end{equation}
where $u=3(1+ \mathsf{w})$ for the barotropic equation of state $p=\mathsf{w}\epsilon$, with the normalization condition
\begin{gather*}
 \sum_{u}B_u=1,
       \label{2.12}
\end{gather*}
where $B_u$ is a contribution of the  u-th component to the energy density at the de Sitter horizon $r_0$. The medium consists of  de Sitter's  vacuum at $u=0$, $\mathsf{w}=-1$. In this case the energy density is $\epsilon=\epsilon_0$.\\
From equation (\ref{2.9}) we can find
\begin{equation}
 {\dot{a}}^2=\frac{8{\pi}G\epsilon{a}^2}{3c^2}- kc^2,  k=0,\pm 1.
       \label{2.13}
\end{equation}
For de Sitter's vacuum $\epsilon=\epsilon_0$ and
\begin{equation}
 \frac{1}{r_0^2}= \frac{8{\pi}G\epsilon_0}{3c^4}=\frac{\Lambda}{3}
       \label{2.14},
\end{equation}
integrating equation (\ref{2.13}), we obtain the following dependences of the scale factor on time and the solution for the inverse function $t(a)$ in the form
\begin{equation}
 a(t)=\left(\frac{ctu\sqrt{B_u}}{2r_0}\right)^{\frac{2}{u}}r_0
       \label{2.15},
\end{equation}
\begin{equation}
 t(a)=\frac{2}{u}\left(\frac{a}{r_0}\right)^{\frac{u}{2}}\frac{r_0}{c\sqrt{B_n}}
       \label{2.16},
\end{equation}
in case $u\neq 0$ and

\begin{equation}
 a(t)=r_0e^{\frac{ct\sqrt{B_0}}{r_0}}
       \label{2.17},
\end{equation}

\begin{equation}
 t(a)=\frac{2}{u}\left(\frac{a}{r_0}\right)^{\frac{u}{2}}\frac{r_0}{c\sqrt{B_n}}
       \label{2.18},
\end{equation}
in case $u=0$.
For the  phantom matter  $\mathsf{w}<-1$ and following equations are  valid:
\begin{equation}
 a(t) \sim |t_0-t|^{\frac{2}{3(1+\mathsf{w})}}
       \label{2.18A},
\end{equation}
where in case $t=t_0$, the Big Rip occurs, and as $t  \rightarrow t_0$ we get cosmological singularity as  $a  \rightarrow \infty$, $ \epsilon  \rightarrow \infty.$

\section{Multicomponent media}

Multicomponent media consisting of different components are investigated. In general, we can consider any multicomponent medium consisting of a set of different components. But not in every case equation (\ref{2.13}) is analytically integrated. Our goal, in this paper, is to find analytic solutions of equation (\ref{2.13}), Therefore we confine ourselves to studying only the cases when (3.6) is integrated analytically in the multicomponent medium under consideration:\\
a) а multicomponent medium consisting of  de Sitter's vacuum, strings, and radiation, $u=0,2,4$.\\
In this case, the energy density, according to (\ref{2.11}),  is expressed as
\begin{equation}
 \epsilon=\epsilon_0\left[B_0+B_2\left(\frac{r_0}{a}\right)^2+B_4\left(\frac{r_0}{a}\right)^4\right]
       \label{2.19},
\end{equation}
and, from equation (\ref{2.13}) we can obtain the following integral, which describes the time dependence of the scale factor $ a$:

\begin{equation}
t=\frac{1}{c}\int\frac{da}{\sqrt{B_0\frac{a^2}{r^2_0}+B_2+B_4\frac{r_0^2}{a^2}}}=\frac{r_0}{2c\sqrt{B_0}}ln\left[2\sqrt{B_0\left(B_0\frac{a^4}{r_0^2}+B_2\frac{a^2}{r_0^2}+B_4\right)} +2B_0\frac{a^2}{r_0^2}+B_2\right]
       \label{2.20},
\end{equation}

b) а multicomponent medium consisting of strings and radiation, $u=2,4$.\\
The energy density and the time dependence on the scale factor are determined 
\begin{equation}
 \epsilon=\epsilon_0\left[B_2\left(\frac{r_0}{a}\right)^2+B_4\left(\frac{r_0}{a}\right)^4\right]
       \label{2.21},
\end{equation}

\begin{equation}
t(a)=\frac{r_0}{2c}\int\frac{dx}{\sqrt{B_2x+B_4}}=\frac{r_0}{c}\sqrt{B_2\frac{a^2}{r_0^2}+B_4}
       \label{2.22},
\end{equation}
 respectively,
 
c) a multicomponent medium consisting of de Sitter's vacuum, dust and ultrastiff matter, $u=0,3,6$.\\
The energy density of a given medium is
\begin{equation}
 \epsilon=\epsilon_0\left(B_0+B_3\left(\frac{r_0}{a}\right)^3+B_6\left(\frac{r_0}{a}\right)^6\right)
       \label{2.23},
\end{equation}
and, from equation (\ref{2.13}), we can obtain the following integral:

\begin{equation}
t=\frac{1}{c}\int\frac{a^2da}{\sqrt{B_0\frac{a^6}{r^6_0}+B_3r_0a^3+B_6r_0^4}}
       \label{2.24}.
\end{equation}

To obtain a dimensionless integral, we introduce the dimensionless quantities:
\begin{gather*}
\left(\frac{a}{r_0}\right)^3=x, a^3=xr_0^3, \\
3a^2da=r_0^3dx
\end{gather*}
under the following conditions:

\begin{gather*}
\Delta=4B_0B_6-B_3^2, \Delta>0, 4B_0B_6>B_3^2;\\
\Delta=0, 4B_0B_6=B_3^2, B_0>0.
\end{gather*}
Then we calculate the integral (\ref{2.24}) and obtain
\begin{equation}
t=\frac{r_0}{3c\sqrt{B_0}}ln\left[2\sqrt{B_0\left(B_0\frac{a^6}{r_0^6}+B_3\frac{a^3}{r_0^3}+B_6\right)} +2B_0\frac{a^3}{r_0^3}+B_6\right]
       \label{2.25},
\end{equation}

d) a multicomponent medium consisting of dust and ultrastiff matter, $u=3,6$.\\
The energy density is
\begin{equation}
 \epsilon=\epsilon_0\left[B_3\left(\frac{r_0}{a}\right)^3+B_6\left(\frac{r_0}{a}\right)^6\right]
       \label{2.26}
\end{equation}

and the integral of  equation (\ref{2.13}) is calculated in the same way as above

\begin{equation}
t(a)=\frac{r_0}{3c}\int\frac{dx}{\sqrt{B_3x+B_6}}=\frac{2r_0}{3c}\sqrt{B_3x+B_6}=\frac{2r_0}{c}\sqrt{B_3\frac{a^3}{r_0^3}+B_6}
       \label{2.27},
\end{equation}

\section{Homogeneous anisotropic models}

	Turning  from homogeneous isotropic models to a more general case that takes anisotropy into account, it should be noted that for the analysis of homogeneous cosmological models Bianchi type classification  is widely used \cite{Bianchi}. The explicit form of the metrics corresponding to various Bianchi types is given in \cite{Petrov}. Without going into detail, we note that the flat and closed Friedmann models considered above are of Bianchi I and IX types respectively.The Kasner metric for empty space, has the form \cite{Kasner}:
\begin{equation}
ds^2=c^2dt^2-a^2(t)d\chi_1^2-b^2(t)d\chi_2^2-c^2(t)d\chi_3^2
       \label{2.28},
\end{equation}
where $a(t)=a_0t^{p_1}$,$b(t)=b_0t^{p_2}$,$c(t)=c_0t^{p_3}$ are the scale factors,  $\chi_\alpha$ are dimensionless coordinates. $a_0$, $b_0$, $c_0$ are dimension constants. $p_\alpha$ are the constant numbers, satisfiying the conditions $(\alpha=1,2,3)$:
\begin{equation}
p_1+p_2+p_3=p_1^2+p_2^2+p_3^2=1
       \label{2.29}
\end{equation}

Their values lie in the intervals:

\begin{equation}
-\frac{1}{3}\leq p_1\leq0, 0\leq p_2\leq \frac{2}{3}, \frac{2}{3}\leq p_3\leq 1.
       \label{2.30}
\end{equation}

Studies of the generalized Kasner solution (for $ p_\alpha $, which depend on spatial coordinates) have shown that the approach to the cosmological singularity $t=0$ has an oscillatory character with a change in the Kasner epochs. However, it should be noted that "although the study of the oscillatory approach to singularity has become an independent branch of mathematical physics \cite{Khalatnikov}",  there is no room for it in the cosmological scenario, since the classical theory for $t \rightarrow 0$ is no longer valid, and it is necessary to investigate quantum solutions.\\

Hawking and Penrose proved general theorems on the inevitability of singularities in general relativity under a number of conditions, in particular, the weak energy condition $p+ \epsilon > 0, \epsilon > 0$, \cite{Hawking}, \cite{Penrose}. These conditions may not be satisfied for some centrally symmetric and cosmological models, in particular for wormholes and the de Sitter model.\\

Now we consider anisotropic cosmological models with  a perfect fluid, taking into account rotation and shear. We will follow the hydrodynamic approach, for matter in the Universe to be considered a perfect fluid. In this case, the following equation is obtained from the  Raychaudhuri   equation (\ref{2.0}):

\begin{equation}
 {\dot{a}}^2=\frac{8{\pi}G\epsilon{a}^2}{3c^2}+\frac{4}{3}\int(\omega^2-\sigma^2)ada- kc^2,  k=0,\pm 1;
       \label{2.31}
\end{equation}

and from the above equation we obtain

\begin{equation}
 t=\int\frac{da}{\sqrt{\frac{8{\pi}G\epsilon{a}^2}{3c^2}+\frac{4}{3}\int(\omega^2-\sigma^2)ada- kc^2}}
       \label{2.32}
\end{equation}
where

\begin{equation}
 \omega^2-\sigma^2=\sum_u\frac{9j_u^2}{u^2B_u^2r_0^2}\left(\frac{r_0}{a}\right)^{2(5-u)}-\frac{\sum_u^2}{a^6}
       \label{2.32А},
\end{equation}

\begin{equation}
 j_u=\frac{J_uc^2}{\epsilon_0r_0^3}
       \label{2.32B}
\end{equation}
and $\sum_u \neq 0$ in case $u=3, 6$.

The divergence of the 4-acceleration $A_{;i}^i$  is zero for the following special cases \cite{Fil'chenkov_02}:
\begin{gather*}
 u=3,6, g^{00} \neq 1; \\
u=0, g^{00}=1
       \label{2.32C}
\end{gather*}

Consider a homogeneous anisotropic multicomponent medium consisting of  de Sitter's vacuum, dust, and stiff  matter, with shear and rotation, i.e., $ \sum_3 \neq 0$, $\sum_6 \neq 0$, $ j_3 \neq 0$, $ j_6 \neq 0$. In this case, from (\ref{2.32}) we can write the following integral:

\begin{equation}
 t=\frac{1}{c}\int\frac{da}{\sqrt{\frac{j_6^2a^4}{12r_0^6c^2}+\frac{B_0a^2}{r_0^2}+\frac{r_0B_3}{a}-\frac{2j_3^2}{3a^2c^2}+\frac{\frac{\sum_3^2}{3c^2}+\frac{\sum_6^2}{3c^2}+B_6}{a^4}}}.
       \label{2.33}
\end{equation}

The following special cases can be derived from the last integral:\\
a) Absence of specific angular momenta $ j_3=j_6=0 $. In this case, calculate the integral (\ref{2.33}) and obtain the following result:

\begin{eqnarray}
 t=\frac{r_0}{3\sqrt{B_0}}ln\left[2\sqrt{B_0\left(B_0\frac{a^6}{r_0^6}+B_3\frac{a^3}{r_0^3}+B_6+\frac{\sum_3^2+\sum_6^2}{3c^2r_0^4}\right)} + 2B_0\frac{a^3}{r_0^3} +B_6+\frac{\sum_3^2+\sum_6^2}{3c^2r_0^4}\right]
\label{2.34}
\end{eqnarray}

\begin{gather*}
\Delta>0, 4B_0B_3>\left(B_6+\frac{\sum_3^2+\sum_6^2}{3c^2r_0^4}\right)^2;\\
\Delta=0, 4B_0B_3=\left(B_6+\frac{\sum_3^2+\sum_6^2}{3c^2r_0^4}\right)^2
 \end{gather*}

b) For large values of scale factor $a$, the integral (\ref{2.33}) is analytically calculated also in the special case as follows 
\begin{equation}
 t=\frac{1}{c}\int\frac{da}{\sqrt{\frac{j_6^2a^4}{12r_0^6c^2}+\frac{B_0a^2}{r_0^2}}}=\frac{j_6}{2B_0c^2}ln\frac{a^2}{r_0^2\sqrt{\frac{a^2}{r_0^2}+\frac{12B_0r_0c^2}{j_6^2}}}
       \label{2.35}
\end{equation}
when $u=0,6$ and $j_6 \neq 0$.

c) For small values of the scale factor a, when $u=6,$ $ j \neq 0$, $\sum_3 \neq 0$, $ \sum_6 \neq 0$, we obtain the integral

\begin{equation}
 t=\frac{1}{c}\int\frac{da}{\sqrt{\frac{\frac{\sum_3^2}{3c^2}+\frac{\sum_6^2}{3c^2}+B_6r_0^4}{a^4}-\frac{2j_3^2}{3a^2c^2}}}=\frac{\sqrt{2}}{3}j_3a^3r_0\frac{1}{\sum_3^2+\sum_6^2+3B_6r_0^4c^2}
       \label{2.36}.
\end{equation}

\section{ Conclusion}
The scale factors time dependences follow either power or exponential laws.  The scale factor dependences on shear are similar to those for stiff matter. The times dependence on scale factors prove to be proportional to rotation.  The scale factor diverges at a finite instant for phantom matter and is time –independent for ekpyrotic matter. It should be noted that all anisotropic cosmological models are considered within the framework of geometric and hydrodynamic approaches. The geometrical approach uses pure geometric classifications according to the Bianchi type. In the hydrodynamical approach, anisotropic cosmological models are  described using the Raychaudhuri equation, and matter in the Universe is considered as a perfect fluid. The Raychaudhuri equation takes into account the expansion, acceleration, rotation and shear of a  perfect  fluid.


\end{document}